\documentclass[10pt,english,conference]{IEEEtran}
\usepackage[T1]{fontenc}
\usepackage[latin9]{inputenc}
\usepackage{units}
\usepackage{amstext}
\usepackage{amssymb}
\usepackage{graphicx}
\usepackage{setspace}
\makeatletter

\author{\IEEEauthorblockN{Jason Cloud and Muriel M\'{e}dard}
\IEEEauthorblockA{Research Laboratory of Electronics\\
Massachusetts Institute of Technology\\
Cambridge, MA USA\\
email: \{jcloud,medard\}@mit.edu}
\and
\IEEEauthorblockN{Douglas Leith}
\IEEEauthorblockA{Hamilton Institute\\
National University of Ireland Maynooth\\
Co. Kildare, Ireland\\
email: doug.leith@nuim.ie}
}

\@ifundefined{showcaptionsetup}{}{%
 \PassOptionsToPackage{caption=false}{subfig}}
\usepackage{subfig}
\makeatother

\usepackage{babel}
\begin{document}

\title{Network Coded TCP (CTCP) Performance over Satellite Networks}
\maketitle
\begin{abstract}
We show preliminary results for the performance of Network Coded TCP
(CTCP) over large latency networks. While CTCP performs very well
in networks with relatively short $RTT$, the slow-start mechanism
currently employed does not adequately fill the available bandwidth
when the $RTT$ is large. Regardless, we show that CTCP still outperforms
current TCP variants (i.e., Cubic TCP and Hybla TCP) for high packet
loss rates (e.g., $>2.5\%$). We then explore the possibility of a
modified congestion control mechanism based off of H-TCP that opens
the congestion window quickly to overcome the challenges of large
latency networks. Preliminary results are provided that show the combination
of network coding with an appropriate congestion control algorithm
can provide gains on the order of 20 times that of existing TCP variants.
Finally, we provide a discussion of the future work needed to increase
CTCP's performance in these networks.\end{abstract}
\begin{IEEEkeywords}
\textit{Network Coding; TCP; High Delay.\vspace{-5pt}}
\end{IEEEkeywords}

\section{Introduction}

\thispagestyle{empty}It is widely known that TCP performs poorly
over satellite networks \cite{caini_transport_2007,pirovano_new_2013}.
The combination of long round-trip times ($RTT$) and high packet
loss rates ($PER$) over these networks create an environment that
seriously degrades the performance of TCP. To overcome the challenges
presented by satellite communication, a large variety of solutions
have been proposed over the years. These range from modifications
to TCP's congestion control algorithm to implementing performance
enhancing proxies (PEPs). Each, of which, usually have their own drawbacks.
In the case of modified TCP protocols, adoption is prevented due to
the specialized nature of the protocol and issues related to fairness
with other TCP variants. In the case of PEPs, increased hardware costs
and issues regarding end-to-end semantics is an issue. In this paper,
we suggest the use of Coded TCP (CTCP) proposed by Kim \textit{et.
al.}, \cite{kim_network_2012}, to overcome a large number of these
issues.

Providing reliable data transport for satellite environments has been
a topic of study since the late 1990's \cite{caini_transport_2007,pirovano_new_2013}.
End-to-end solutions typically involve tuning TCP so that the long
$RTT$s representative of satellite links do not negatively impact
performance. Two versions that perform well over satellite networks
are TCP Cubic \cite{ha_cubic:_2008} and TCP Hybla \cite{caini_tcp_2004}.
Cubic, designed for high speed networks, and Hybla, designed for heterogeneous
networks, use a congestion window algorithm that increases the congestion
window size ($cwnd$) independently from the $RTT$. This makes either
version useful in environments with high delay. Unlike TCP Cubic,
Hybla was developed to also reduce the impact of multiple losses,
inappropriate timeouts, and burstiness making it a more logical option
for use over satellite links. When compared with each other, studies
have shown that Hybla performs better than Cubic under high PERs while
the reverse is true under low $PER$'s \cite{trivedi_comparative_2010}.
Regardless, both experience performance degradation under high losses.
A more recent protocol, Loss-Tolerant TCP (LT-TCP) \cite{ganguly_loss-tolerant_2012,sharma_mplot:_2008},
combines Reed-Solomon (RS) coding with TCP to overcome this issue,
but it requires the use of explicit congestion control (ECN) and the
RS code can result in performance loss due to decoding errors. CTCP
circumvents these issues by using a congestion control algorithm that
does not rely on feedback from lower layers and network coding eliminates
the possibility of decoding errors while helping to overcome packet
losses.

In lieu of changes to TCP, PEPs are another common method used to
increase performance over satellite links. A TCP flow is generally
terminated at the gateway to the satellite link, a protocol specifically
designed for the satellite system (usually one that is proprietary)
is used to transmit data over the satellite network, and a new TCP
session is setup on the other side of the satellite link to complete
the connection. This implementation poses two issues. First, the cost
of implementing a PEP at the satellite network gateway may be high.
Second, the termination of TCP sessions at the PEP violates end-to-end
semantics such as IPSEC \cite{pirovano_new_2013}. Again, CTCP has
the potential to eliminate the need for PEPs while providing the same
level of service over satellite links.

In this paper, we explore CTCP's performance in high $RTT$ environments
to determine if we can provide resilience in the presence of high
packet loss rates, in addition to achieving the performance of TCP
Cubic or TCP Hybla under no packet losses. We first provide a description
in Section \ref{sec:CTCP-Overview} of the existing, well tested version
of CTCP that uses a TCP Reno style slow-start mechanism. This version
is designed to provide robustness to packet losses through the use
of network coding, but the congestion window management is ill-suited
to large bandwidth-delay products (BDP). In Section \ref{sec:CTCP-Long-RTT}
we measure the performance of a modified version of CTCP that opens
$cwnd$ in a manner similar to H-TCP \cite{leith_h-tcp:_2004} to
show that it is indeed possible to achieve high performance with large
$PER$'s and $RTT$'s. Finally, we conclude in Section \ref{sec:Future-Work}
by proposing areas of possible future work.

\section{CTCP Overview\label{sec:CTCP-Overview}}

The development of CTCP \cite{kim_network_2012} has shown how the
integration of network coding with TCP can provide significant benefits
over existing TCP variants, especially in high packet loss environments.
These gains are a direct result of the combination of both network
coding and CTCP's congestion window management. The remainder of this
section will provide a brief introduction into both of these mechanisms.
For a more detailed explanation of CTCP's implementation and performance,
the reader should refer to \cite{kim_network_2012}.

\subsection{Network Coding}

One of the key features of CTCP is the use of network coding to aid
in recovery from packet losses and the capability to decrease overhead
by limiting the number of required retransmissions. The gains provided
by network coding are twofold: network coded packets can be used to
provide forward error correction in the case of lost packets, and
also simplifies feedback and retransmissions (should they be needed).
In its current implementation, CTCP uses a systematic random linear
code \cite{ho_random_2006}. As an example, consider the transfer
of packets $p_{1}\ldots p_{k}$ between a server and client. Each
packet $p_{i},i\in\{1,\ldots,k\}$ is first sent uncoded followed
by a number of network coded packets where every coded packet $c_{i}$
is a random linear combination of the packets $p_{1},\ldots,p_{k}$,
i.e., 
\begin{equation}
c_{i}=\sum_{j=1}^{k}\alpha_{j}p_{j},
\end{equation}
each $\alpha_{j}\in\mathbb{F}_{2^{q}}$ is randomly chosen, and $q$
is large enough to ensure linear independence among all network coded
packets with high probability (the current implementation draws $\alpha_{j}$
from $\mathbb{F}_{256}$). Should retransmissions be needed, additional
network coded packets are generated and sent to the client. The number
of coded packets sent along with the uncoded packets is dynamically
determined based on an estimate of the path's packet loss probability,
while the number of packets sent as a result of feedback is determined
by both the number degrees of freedom ($dofs$) required by the client
to decode and the estimated packet loss probability.

\subsection{Congestion Control}

The second feature of CTCP that is a major contributor to the observed
gains is the congestion window management. CTCP uses a modified version
of TCP's Additive Increase, Multiplicative Decrease (AIMD) algorithm
that was designed to be compatible with network coding. Specifically,
the current implementation modifies the multiplicative back-off factor,
$\beta$, to be 
\begin{equation}
\beta=\frac{RTT_{\min}}{RTT},
\end{equation}
where $RTT_{\min}$ is the path's estimated true round-trip propagation
delay (which is assumed to be the lowest per-packet $RTT$ observed
during the lifetime of a connection) and $RTT$ is the last measured
round-trip time. The congestion window is increased using TCP Reno's
increase mechanism (i.e., the slow-start mode increases $cwnd$ by
1 for every received acknowledgement, otherwise $cwnd$ is increased
by $\nicefrac{1}{cwnd}$).

This approach, in effect, assumes that the increase of a packet's
$RTT$ is solely due to the queuing of packets along the path, which
is an indication of congestion. If a packet is lost at random and
$RTT=RTT_{\min}$ (i.e., it is not lost due to congestion), then $cwnd$
is not reduced. On the other hand if $RTT>RTT_{\min}$, a packet loss
is interpreted as congestion and $cwnd$ is reduced by a factor of
$\beta$. While this approach does a fairly good job at distinguishing
between packet losses due to congestion and poor channels, it has
a few characteristics that may not work well for satellite networks.
A more detailed discussion is provided in Section \ref{sec:Future-Work}.

\subsection{Performance over Short $RTT$ Networks}

Using the mechanisms summarized above, \cite{kim_network_2012} implemented
CTCP as a SOCKSv5 proxy in user space and measured its performance
over a wide range of conditions (although all measurements were made
using round-trip times representative of terrestrial networks). Kim\textit{
et. al. }showed that CTCP can achieve goodput efficiencies greater
than 90\% for packet loss rates less than 0.2 while the performance
of standard TCP variants is severely impacted. Another important aspect
of CTCP is that it is friendly/fair with standard TCP, unlike some
TCP variants that work well over satellite networks but are unfriendly
to other TCP variants (e.g., TCP Hybla and Cubic TCP \cite{urke_tcp_2012}).
This is important since we are interested in providing an end-to-end
solution. Therefore, we would like to ensure that if the bottleneck
link is not the satellite link, CTCP does not adversely impact the
performance of TCP flows not traversing the satellite.

\section{CTCP Performance in Satellite Networks\label{sec:CTCP-Long-RTT}}

While previous results show that CTCP has great potential in networks
with high packet losses and low $RTT$, no measurements exist for
networks with large $RTT$. This section will explore the potential
for CTCP to provide improved performance in environments with large
delays using a testbed located at the Hamilton Institute, NUI Maynooth,
Ireland.

\begin{figure}
\begin{centering}
\includegraphics[width=0.7\columnwidth]{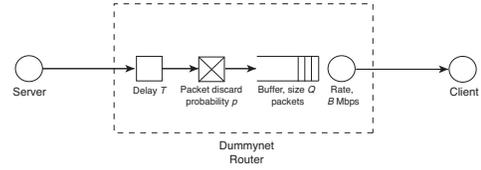}\vspace{-5pt}
\par\end{centering}

\caption{Schematic of experimental testbed.\vspace{-15pt}\label{fig:Testbed-Schematic}}
\end{figure}

The testbed used to collect measurements consists of commodity servers
(Dell Poweredge 850, 3GHz Xeon, Intel 82571EB Gigabit NIC) connected
via a router and gigabit switches. A diagram of the setup is shown
in Figure \ref{fig:Testbed-Schematic}. Sender and receiver machines
used in the tests both run a Linux 2.6.32.27 kernel. The router is
a commodity server running FreeBSD 4.11 and \texttt{ipfw-dummynet}.
Data is transferred between the sender and receiver machines using
\texttt{rsync} (version 3.0.4) and the appropriate TCP version.

\begin{figure*}
\begin{centering}
\vspace{-5pt}\subfloat[$RTT=500\text{ ms}$]{\includegraphics[width=0.5\textwidth]{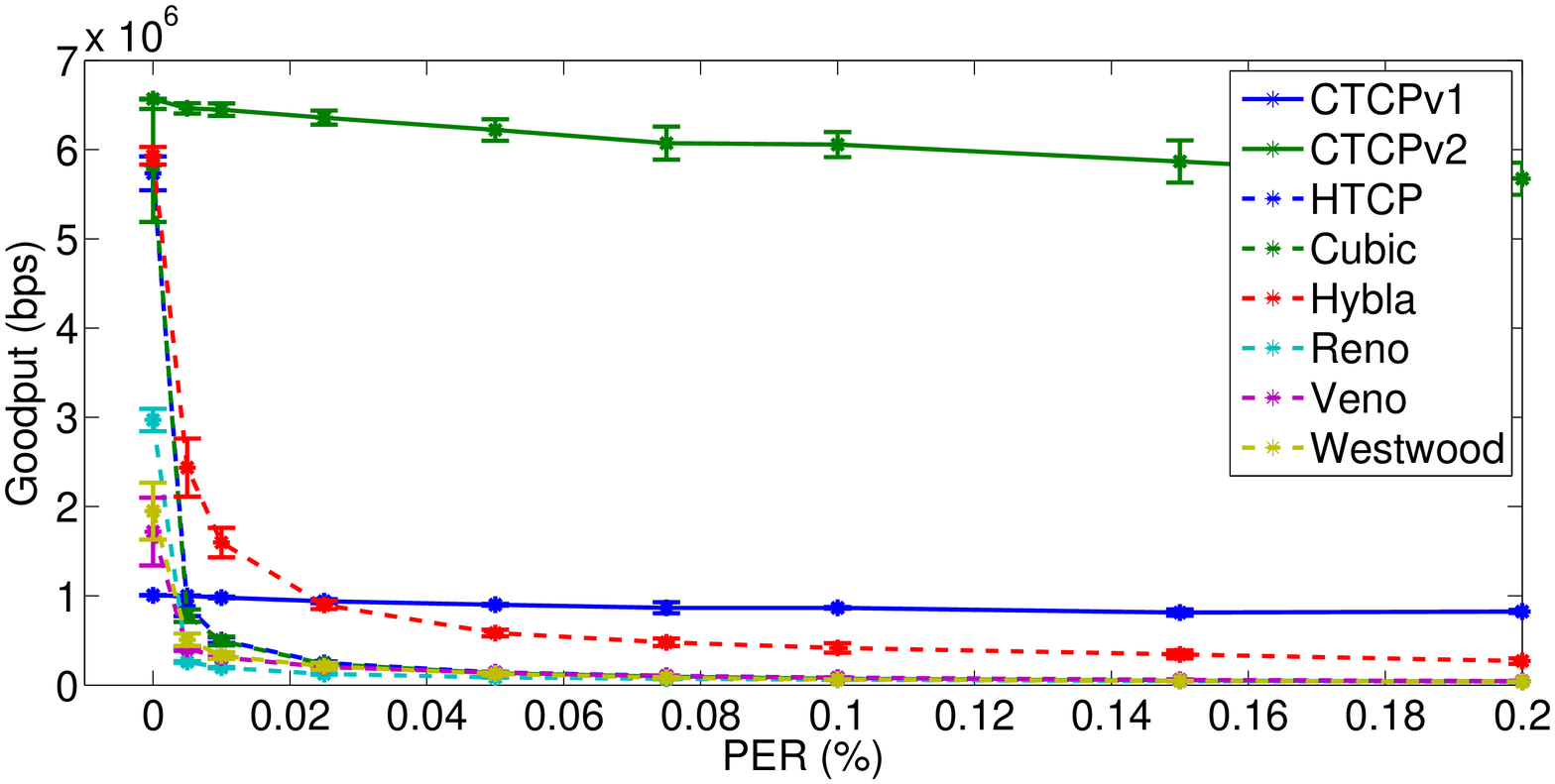}

}\subfloat[$RTT=600\text{ ms}$]{\includegraphics[width=0.5\textwidth]{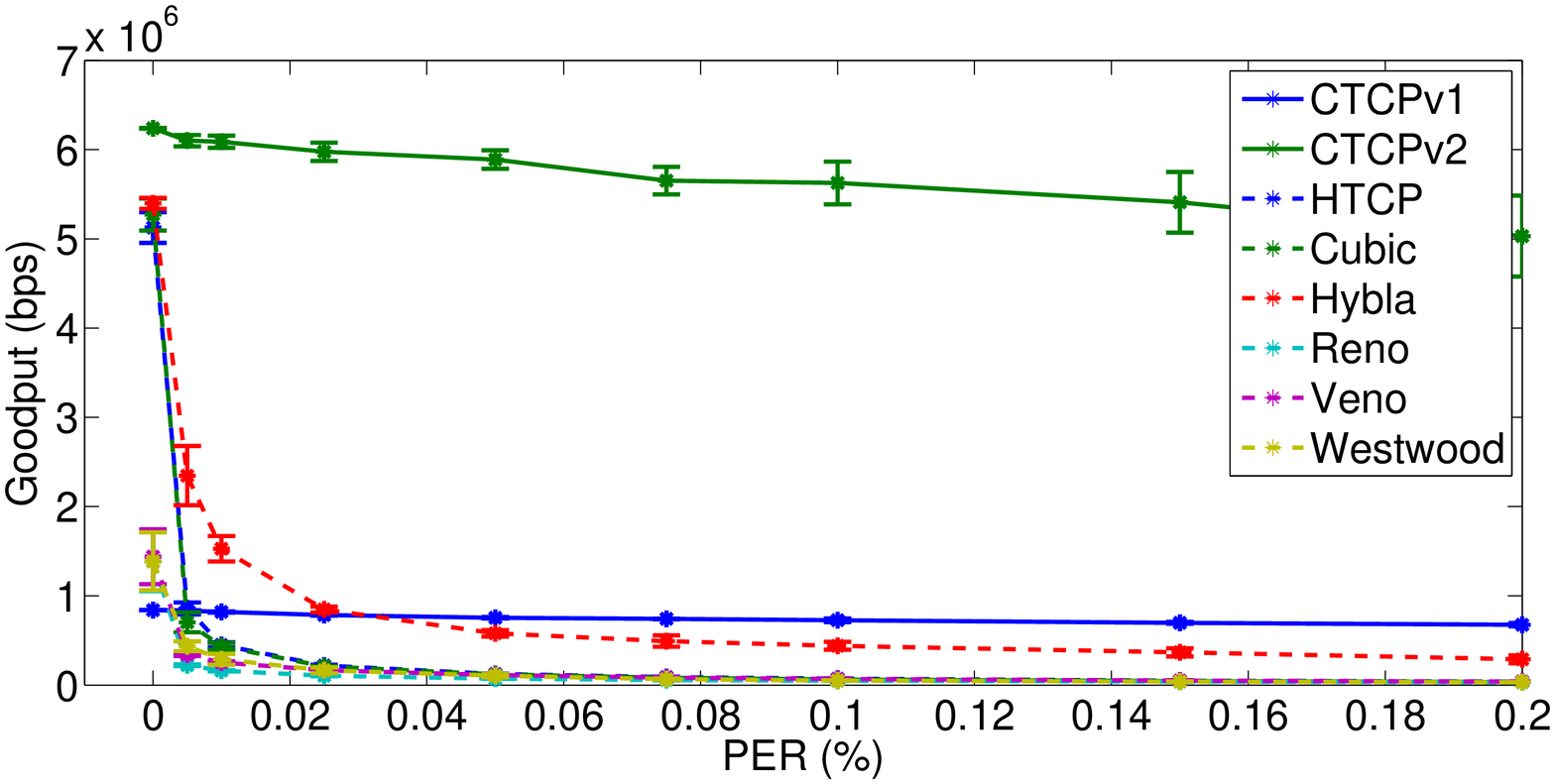}

}\vspace{-10pt}
\par\end{centering}

\begin{centering}
\subfloat[$RTT=700\text{ ms}$]{\includegraphics[width=0.5\textwidth]{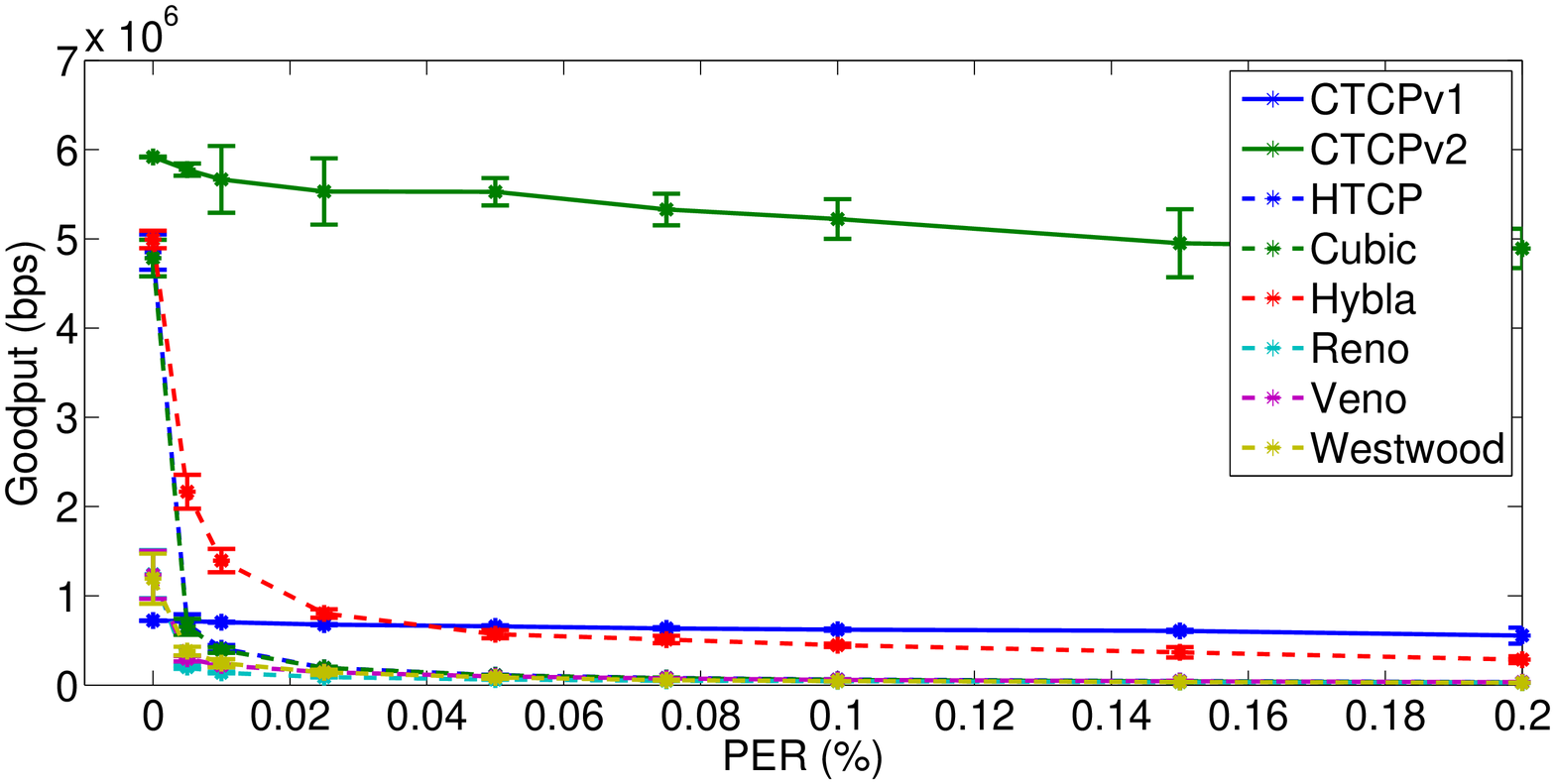}

}\subfloat[$RTT=800\text{ ms}$]{\includegraphics[width=0.5\textwidth]{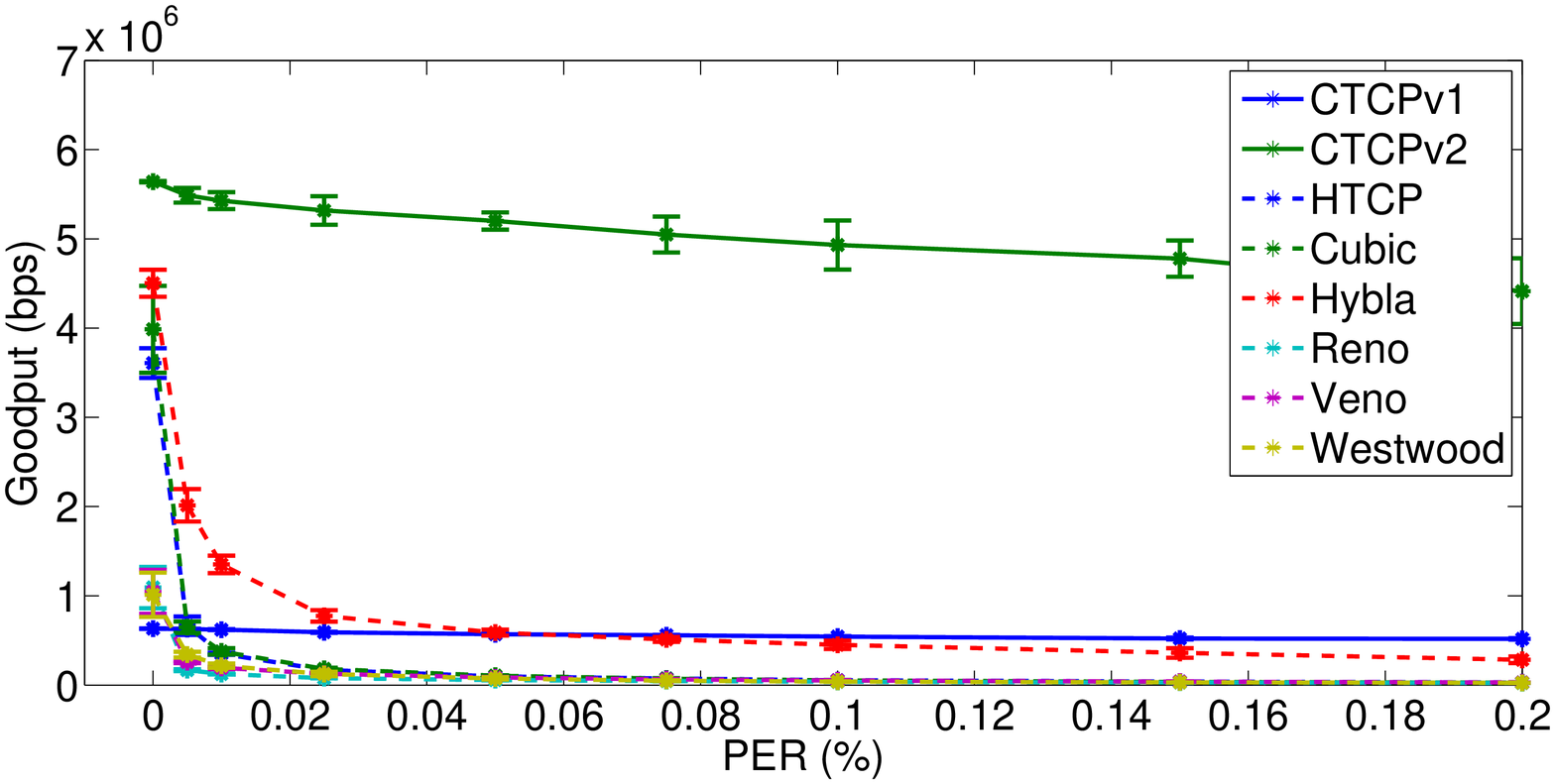}

}
\par\end{centering}

\caption{Comparison of TCP variants with varying $PER$ and $RTT$ with a link
rate of 10 Mbps. Each bar shows the mean goodput, while the error
bars show one standard deviation. \vspace{-15pt}\label{fig:Comparison-of-TCP}}
\end{figure*}

Each version of TCP, other than CTCP, is implemented within the kernel
making it easy to select the appropriate variant. In the case of CTCP,
it is implemented in user space as a SOCKSv5 proxy with the forward
proxy located on the client and the reverse proxy located on the server.
Traffic between the proxies is sent using CTCP. Therefore, a client's
request is first directed to the local forward proxy, transmitted
to the reverse proxy, and then forwarded to the appropriate port on
the server. The server responds using the reverse process. In order
to use \texttt{rsync}, \texttt{proxychains} (version 3.1) was used
to direct traffic to the proxy.

A series of tests were conducted using the following TCP variants:
CTCP, Cubic, Hybla, Reno, Veno, and Westwood. A 20 MB file download
is used, along with varying $PER$'s and $RTT$'s, to characterize
the performance of each TCP version. Figure \ref{fig:Comparison-of-TCP}
provides a summary of each version's mean goodput as a function of
the $PER$. Each test was run a minimum of three times and a maximum
of ten times depending on the amount of time need to complete the
20 MB download.

The performance of CTCP, labeled ``CTCP v1'' in Figure \ref{fig:Comparison-of-TCP},
in networks with large $RTT$ and low $PER$ is significantly poorer
than two of the TCP variants designed for these network types (i.e.,
TCP Hybla and TCP Cubic). For $PER$ greater than $2.5\%$, CTCP begins
to outperform both of these TCP variants for most of the $RTT$s tested.
In fact, the goodput of CTCP remains relatively constant as the $PER$
increases to $20\%$ while the goodput of the other TCP variants approaches
zero quickly.

The additive increase portion of CTCP's current congestion control
algorithm is the primary reason for its poor performance at low $PER$.
To overcome the challenges related to long $RTT$'s, a modified version
of CTCP, labeled ``CTCP v2'' in Figure \ref{fig:Comparison-of-TCP},
was implemented that increases $cwnd$ in a manner consistent with
H-TCP (see \cite{leith_h-tcp:_2004} for more details). Because $cwnd$
is no longer dependent on the $RTT$, it can increase rapidly allowing
it to use the available capacity more efficiently. In addition to
the use of network coding and the unmodified multiplicative $cwnd$
back-off approach, CTCP is able to maintain a large throughput for
$PER$'s as high as $20\%$. In fact, measurements indicate that this
modified version of CTCP provides a gain of approximately 21 times
that of TCP Hybla for a $PER$ of $20\%$ and $RTT$ of $500$ ms
over a link with a bandwidth of 10 Mbps.

\begin{figure}
\begin{centering}
\vspace{-15pt}\subfloat[$RTT=500$ ms]{\includegraphics[width=0.45\columnwidth]{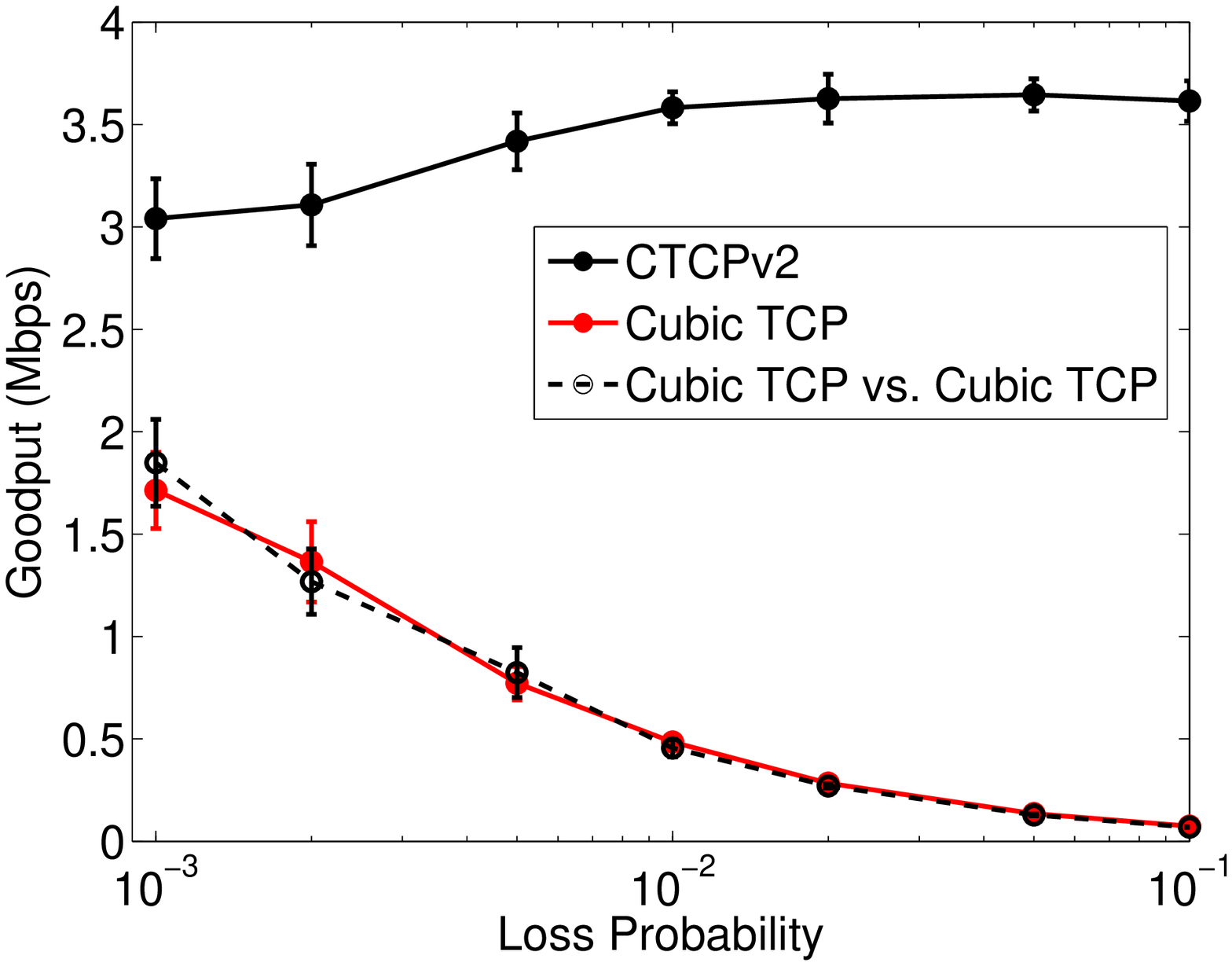}

}\subfloat[$RTT=800$ ms ]{\includegraphics[width=0.45\columnwidth]{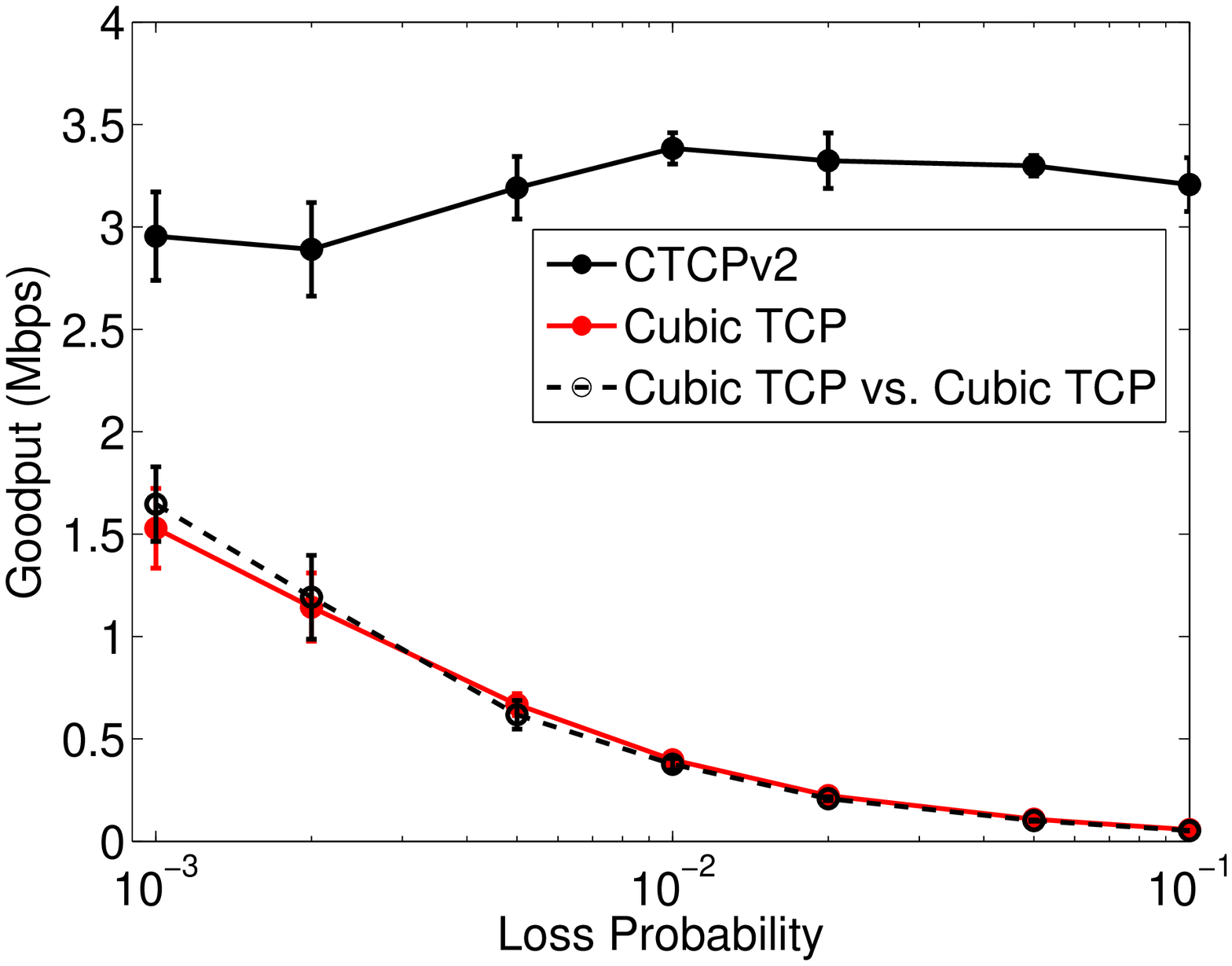}

}
\par\end{centering}

\caption{Goodput versus packet loss rate for (i) a Cubic TCP flow and CTCPv2
flow sharing a link (solid lines), and (ii) two Cubic TCP flows sharing
a link (dashed line). The link rate is 5 Mbps and the error bars show
one standard deviation. \vspace{-25pt}\label{fig:fairness_modCTCP}}

\end{figure}

In addition, preliminary testing has shown that this version of CTCP
is friendly with existing TCP versions. Each sub-figure in Figure
\ref{fig:fairness_modCTCP} provides a comparison of the throughput
for two tests. The dotted line labeled ``Cubic TCP vs. Cubic TCP''
shows the throughput obtained by one Cubic TCP flow competing against
a second Cubic TCP flow. The solid lines show the second test where
a Cubic TCP flow is competing against a CTCP flow. The indication
of fairness is provided by the similarity of the solid Cubic TCP line
and the dotted line.

\begin{figure}
\vspace{-10pt}\subfloat[$PER=0.5\%$]{\includegraphics[width=0.45\columnwidth]{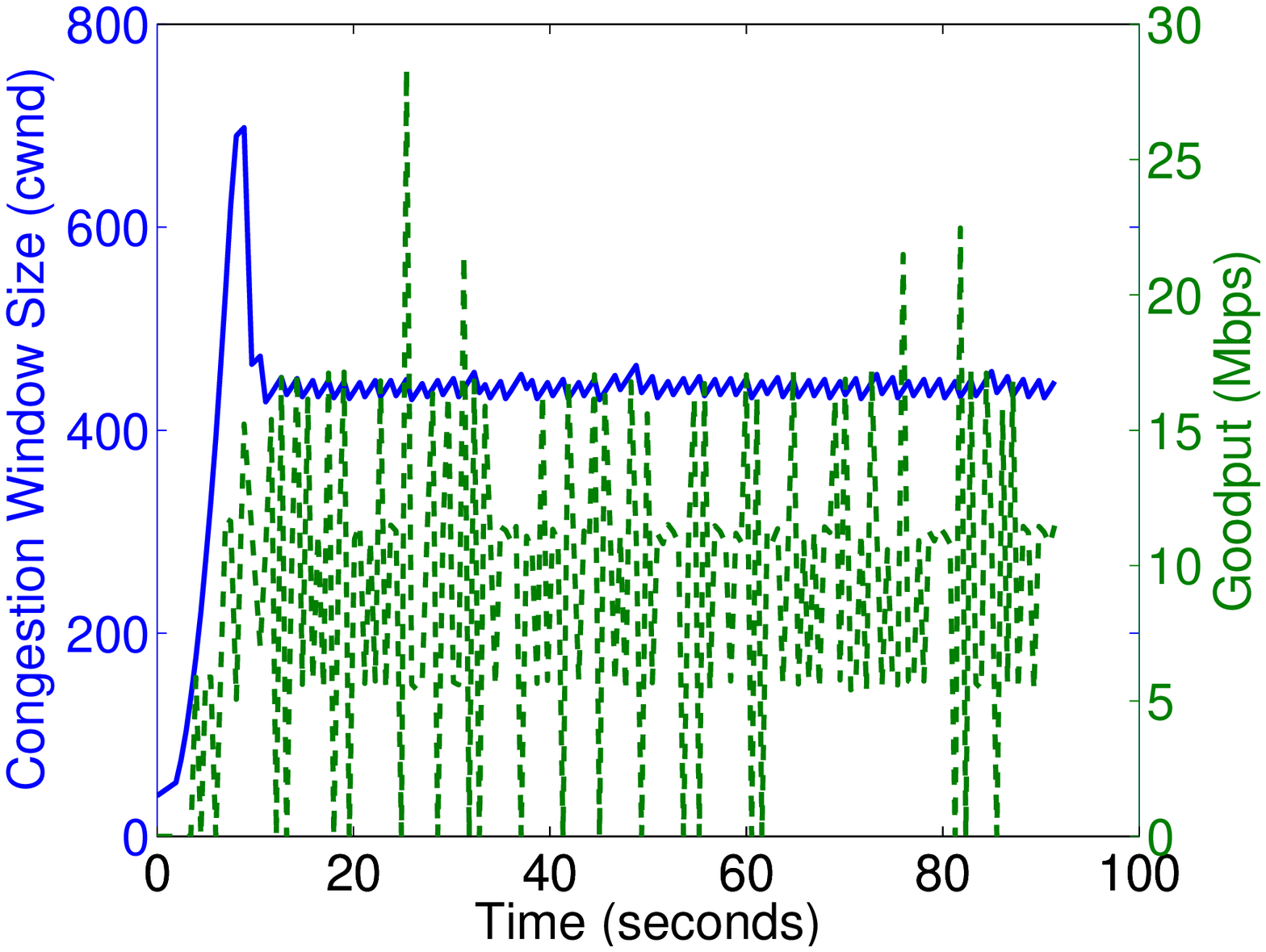}

}\subfloat[$PER=20\%$]{\includegraphics[width=0.45\columnwidth]{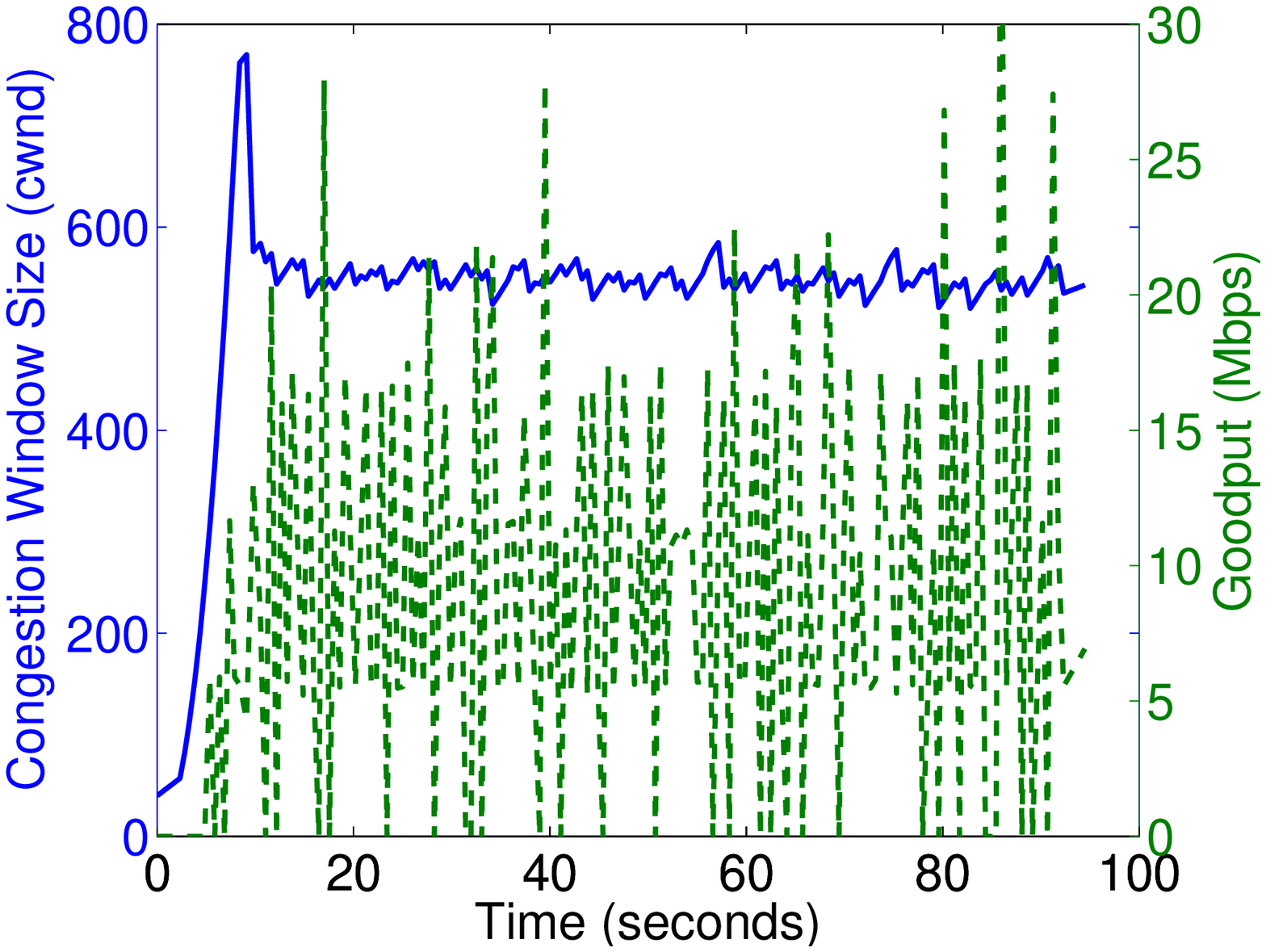}

}

\caption{Trace of CTCPv2 over a 10 Mbps link with a $RTT$ of 500 ms. The dotted
line shows the the instantaneous goodput and the solid line shows
$cwnd$. The mean goodput in (a) and (b) is 9.19 Mbps and 8.92 Mbps
respectively.\vspace{-17pt}\label{fig:Trace-of-CTCPv2}}
\end{figure}

While these results are promising, additional work is required. This
is evident in the trace of the goodput and $cwnd$ shown in Figure
\ref{fig:Trace-of-CTCPv2}. The instantaneous goodput is highly variable,
which causes delay jitter as packets are delivered to higher layers.
Possible causes of this may be an underestimate of the packet loss
probability or an underestimate of the number of coded packets needed.
Either case creates the distinct decode events shown in the figure.
Regardless, the potential for greatly increasing performance at the
transport layer is evident. Not only can throughput performance be
drastically improved, but this solution appears, at first glance,
to be backward compatible with existing TCP variants.

\section{Conclusions and Future Work\label{sec:Future-Work}}

In Section \ref{sec:CTCP-Overview}, an overview of CTCP is presented
and a brief overview of CTCP's performance in networks with short
$RTT$ was provided. In Section \ref{sec:CTCP-Long-RTT}, CTCP's performance
is compared with other TCP variants in networks with $RTT$'s similar
to those that would be observed with satellite communications. Measurements
showed that the current implementation of CTCP performed worse than
some existing TCP versions for small $PER$, but outperformed other
TCP versions for high $PER$. One of the primary causes of this was
discussed and an alternate $cwnd$ increase mechanism is used to highlight
that some minor changes to the current congestion control algorithm
can significantly improve performance in these environments. While
previous sections discussed several areas of future research such
as this, the remainder of this section will introduce additional future
research directions.

Additional research into CTCP's congestion control algorithm is required.
First, the current implementation relies on the $RTT$ of the path
in order to increase $cwnd$. For connections with large $RTT$'s,
this is obviously an issue. More aggressive methods for increasing
$cwnd$, such as using a H-TCP like mechanism, need to be thoroughly
developed while still maintaining interoperability with network coding
and fairness with legacy TCP variants. Second, the method for determining
whether a packet is lost due to congestion or due to a poor link is
also an issue. The multiplicative back off method currently used (i.e.,
$\beta=\nicefrac{RTT_{\min}}{RTT}$) works well when jitter in the
$RTT$ measurement is primarily caused by the filling of queues. It
fails to work properly when the delay jitter is caused by something
else. For example, if the delay jitter is caused by a particular medium-access
(MAC) method, it is likely that $\beta<1$ causing $cwnd$ to collapse.
This was observed in measurements taken over a WiMax network \cite{kim_network_2012}
where the MAC's scheduling algorithm caused large variations in the
$RTT$. Methods using feedback from the network, such as LT-TCP, which
uses explicit congestion notification (ECN) \cite{ganguly_loss-tolerant_2012},
are possibilities although we would like to ensure that CTCP operates
irrespective of lower layer implementations.

The use of network coding in CTCP is critical for overcoming packet
losses and providing high throughput, but little is understood about
how to adjust the number of packets coded together (i.e., the coding
window). The current implementation of CTCP uses a fixed size block,
or generation, scheme for generating network coded packets. This is
not optimal given the user's requirements since there is an inherent
tradeoff between throughput and delay as the block, or generation,
size is changed \cite{zeng_joint_2012}. Dynamically adjusting the
block size, or using a sliding coding window approach, to meet the
user's throughput/delay requirements is also a topic of ongoing research.
Furthermore, the interaction between congestion avoidance and network
coding is not fully understood. The current implementation treats
both congestion avoidance and network coding separately, yet there
is evidence that intelligently merging the two can provide a performance
increase.

In summary, CTCP has potential to greatly improve network performance
over existing transport layer protocols in the presence of both high
packet error rates and round-trip times. Initial measurements have
shown significant gains in goodput over existing TCP versions, but
additional research is needed to tune both the congestion control
algorithm and network coding parameters to ensure proper functionality.

\section*{Acknowledgment}

This work is sponsored, in part, by the Assistant Secretary of Defense
(ASD R\&E) under Air Force Contract \# FA8721-05-C-0002.  Opinions,
interpretations, recommendations and conclusions are those of the
authors and are not necessarily endorsed by the United States Government.

\bibliographystyle{IEEEtran}
\bibliography{CTCP_Performance_over_Satellite_Networks}

\end{document}